\documentclass[a4paper,11pt]{article}
\usepackage{pos}
\usepackage{siunitx}
\usepackage{amsmath}
\usepackage[nameinlink]{cleveref}
\crefname{equation}{Eq.}{Eqs.}

\DeclareSIUnit{\parsec}{\mathrm{pc}}
\DeclareSIUnit{\gauss}{\mathrm{G}}
\DeclareSIUnit{\erg}{\mathrm{erg}}
\DeclareSIUnit{\year}{\mathrm{yr}}

\title{A Bayesian Framework for UHECR Source Association and Parameter Inference}

\author*[a]{Keito Watanabe}
\author[b,d]{Anatoli Fedynitch}
\author[c]{Francesca Capel}
\author[d]{Hiroyuki Sagawa}

\affiliation[a]{Institute for Astroparticle Physics, Karlsruhe Institute of Technology, Hermann-von-Helmholtz-Platz 1, 76344 Eggenstein-Leopoldshafen, Germany}

\affiliation[b]{Institute of Physics, Academia Sinica, No. 128, Sec. 2, Academia Rd, Nangang District, 11529 Taipei, Taiwan}

\affiliation[c]{Max Planck Institute for Physics, Boltmannstr. 8, Garching 85748, Germany}

\affiliation[d]{Institute for Cosmic Ray Research, the University of Tokyo, 5-1-5 Kashiwa-no-ha, Kashiwa, Chiba 277-8582, Japan}

\emailAdd{keito.watanabe@kit.edu}
\emailAdd{anatoli@gate.sinica.edu.tw}
\emailAdd{capel@mpp.mpg.de}
\emailAdd{hsagawa@icrr.u-tokyo.ac.jp}

\abstract{
    The identification of potential sources of ultra-high-energy cosmic rays (UHECRs) remains challenging due to magnetic deflections and propagation losses, which are particularly strong for nuclei. In previous iterations of this work, we proposed an approach for UHECR astronomy based on Bayesian inference through explicit modelling of propagation and magnetic deflection effects. The event-by-event mass information is expected to provide tighter constraints on these parameters and to help identify unknown sources. However, the measurements of the average mass through observations from the surface detectors at the Pierre Auger Observatory already indicate that the UHECR masses are well represented through its statistical average. In this contribution, we present our framework which uses energy and mass moments of $\ln A$ to infer the source parameters of UHECRs, including the mass composition at the source. We demonstrate the performance of our model using simulated datasets based on the Pierre Auger Observatory and Telescope Array Project. Our model can be readily applied to currently available data, and we discuss the implications of our results for UHECR source identification.
}

\ConferenceLogo{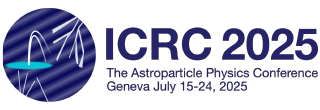}

\FullConference{39th International Cosmic Ray Conference (ICRC2025)\\
 15–24 July 2025\\
Geneva, Switzerland\\}

\begin{document}
\maketitle

\section{Introduction}

With more than 20 years of observations, the Pierre Auger Observatory (Auger) and Telescope Array Project (TA) have provided us with a wealth of data on ultra-high energy cosmic rays (UHECRs), and will provide us with even more data with the AugerPrime upgrade \cite{Castellina:2019irv} and TAx4 experiment \cite{TelescopeArray:2021dri}. However, the association of UHECRs with their potential sources remains a challenge due to the limited statistics and the complex propagation effects in the intergalactic and Galactic medium. In particular, the mass composition of UHECRs has not been measured on an event-by-event basis, but rather only as a statistical average of all observed UHECRs. This limits the constraints placed on the rigidity-dependent deflections and mass-dependent propagation losses of each UHECR. Nevertheless, with sophisticated analyses techniques, the combination of currently available arrival directions, energies, and mass composition data can be used to constrain the properties of potential sources and the propagation effects of UHECRs \cite{PierreAuger:2023htc}. \par 

In previous works, we developed a Bayesian hierarchical model capable of inferring ultra-high-energy cosmic ray (UHECR) source parameters, such as luminosity and spectral indices, while accounting for mass-dependent propagation effects using the nuclear propagation code \texttt{PriNCe}~\cite{Watanabe:2023fdz,Watanabe:2023bmf,Watanabe:2025hsj,Heinze:2019jou}. However, earlier iterations of this model required assumptions about the arrival composition falling within a predefined mass group, limiting its direct applicability to observational data. In this work, we relax this assumption by explicitly incorporating the mean and variance of the logarithmic mass ($\ln A$) distribution measured at Earth. We demonstrate that combining these mass composition moments with individual event energies is sufficient to robustly infer source parameters, including the initial UHECR mass composition at the sources. We present detailed analyses of simulated datasets reflecting the exposures of the Pierre Auger Observatory and the Telescope Array, highlighting the method's performance and applicability to currently available observational data.


\section{Model}

As with our previous work, the model used in this work is based on a Bayesian hierarchical framework, where we set priors on the source parameters, such as the luminosity and spectral indices, as well as the mass fractions of UHECRs at the source. The inference is performed using the probabilistic programming language \texttt{stan} \cite{stan:2024}, which performs Markov Chain Monte Carlo (MCMC) sampling through the Hamiltonian Monte Carlo (HMC) algorithm to directly obtain the posterior distributions of the model parameters. \par 

\subsection{Modelling UHECR Propagation with \texttt{PriNCe}}

To describe the mass-dependent propagation effects, we use the nuclear propagation code \texttt{PriNCe} \cite{Heinze:2019jou}. This code describes its nuclear propagation as a series of partial differential equations and numerically solves for the co-moving density of UHECRs $Y_i(E_i, z) = N_i(E_i, z) / (1 + z)^3$ for each mass component $i$ at a given energy $E_i$ and redshift $z$. The propagation equation is given by:

\begin{equation}
    \partial_z Y_i(E_i, z) = -\partial_E(b_\mathrm{ad} Y_i) - \partial_E(b_{e^+ e^-} Y_i) - \Gamma_i Y_i + \sum_j Q_{i \rightarrow j} Y_j + J_i,
    \label{eq:prince_propa_eq}
\end{equation}
which describes (in order from left to right) the adiabatic losses, pair production losses and photo-nuclear interactions (decay and re-injection). The last term $J_i = J_i(E_i, z, A^S_i)$ describes the injection from a distribution of homogeneously distributed and isotropically emitting sources, where $A^S_i$ denotes the mass number of component $i$ at the source. In this work, we describe the photo-nuclear and photo-hadronic cross sections with \texttt{TALYS} \cite{Koning:2005ezu} and \texttt{SOPHIA} \cite{Mucke:1999yb} respectively. \par 

\noindent \textbf{Source model:} While the same propagation model is used to describe the UHECRs emitted from the source and from background sources, we model them separately. Given $N_S$ sources, each source (labelled as $k$) with a given luminosity $L_k$ at distance $d_k$ (flux $F_k \propto L_k / 4\pi d_k^2$) is treated as a point source that emits UHECRs with a single injection mass composition $A^S_l$ isotropically without any background sources. This amounts to setting $J_i = 0$ in \Cref{eq:prince_propa_eq} and setting the initial condition of the co-moving density $Y_i$ at $z_k = z(d_k)$ with the energy distribution of UHECRs for a single injected mass at the source (i.e. the source spectrum). The source spectrum follows that of \cite{PierreAuger:2016use}: 
\begin{equation}
    \mathcal{J}^S_{lk}(E_i, Z^S_l \: | \: \alpha_k, R_\mathrm{max}) \propto \left(\dfrac{E_i}{\SI{1}{\exa\electronvolt}}\right)^{-\alpha_k} \times 
        \begin{cases}
            1 & : E_i < Z^S_l R_\mathrm{max} \\
            \exp\left(1-\dfrac{E_i}{Z^S_l R_\mathrm{max}}\right) & : E_i \geq Z^S_l R_\mathrm{max}
        \end{cases}
    \label{eq:source_model}
\end{equation}
where $Z^S_l$ denotes the charge number corresponding to the injected mass $A^S_l$. Each source is characterized by a spectral index $\alpha_k$ and a maximal rigidity $R_\mathrm{max}$, which is assumed to be identical for all injected elements. Although previous studies have indicated that maximal rigidity may vary between sources due to differences in their acceleration properties, we currently fix $R_\mathrm{max} = \SI{1.7}{\exa\electronvolt}$ for all sources following earlier analyses~\cite{Heinze:2019jou,Ehlert:2022jmy}. The normalization constant, representing the number of injected particles with mass $A^S_l$ per comoving volume, per unit energy, and per year, is determined within the forward model implemented in \texttt{stan}.
\par 

\noindent \textbf{Background Model: } 
Our background model assumes the continuous injection of UHECRs from sources homogeneously distributed from redshift $z_\mathrm{max} = 3$ down to a truncation redshift $z_\mathrm{trunc}$. This is modeled by setting initial conditions $Y_i(E_i, z) = 0$ and defining the injection function $J_{il}$ as:
\begin{equation}
    J_{il}(E_i, z, Z^S_l \mid \alpha_0, R_\mathrm{max}) = 
    \begin{cases}
        n_\mathrm{evol}(z)\,\mathcal{J}^S_{lk}(E_i, Z^S_l \mid \alpha_0, R_\mathrm{max}), & z \geq z_\mathrm{trunc} \\[0.5em]
        0, & z < z_\mathrm{trunc}.
    \end{cases}
    \label{eq:background_model}
\end{equation}
Analogous to the source model, the background model is described by a spectral index $\alpha_0$, a fixed maximal rigidity $R_\mathrm{max} = \SI{1.7}{\exa\electronvolt}$, mass fractions $f^A_{l0}$ for each injected composition, and a total normalization flux $F_0$. We assume that the source density $n_\mathrm{evol}(z)$ evolves with redshift according to the star formation rate (SFR) parameterization from Ref.~\cite{Yuksel:2008cu}.

Using \Crefrange{eq:source_model}{eq:background_model}, we generate a database of solutions by solving \Cref{eq:prince_propa_eq} on a grid of spectral indices and injection masses. In our analysis, we consider $N^A_\mathrm{inj} = 5$ injected mass groups: proton ($A = 1$), helium ($A = 4$), nitrogen ($A = 14$), silicon ($A = 28$), and iron ($A = 56$). The resulting comoving density at Earth, $Y_{ilk}^\mathrm{Earth} := Y_i(E_i, z=0, Z^S_l \mid \alpha_k)$, is used to calculate the flux at Earth $\mathcal{J}^\mathrm{Earth}_{lk}$, as well as the spectrum-weighted mean and variance of $\ln A$. These quantities are obtained by summing over all mass species at Earth:

\begin{equation}
    \mathcal{J}^\mathrm{Earth}_{lk} = \sum_i Y_{ilk}^\mathrm{Earth}, \quad 
    \langle \ln A \rangle_{lk} = \frac{\sum_i Y_{ilk}^\mathrm{Earth} \ln A_i}{\sum_i Y_{ilk}^\mathrm{Earth}}, \quad 
    \mathrm{Var}(\ln A)_{lk} = \frac{\sum_i Y_{ilk}^\mathrm{Earth} (\ln A_i)^2}{\sum_i Y_{ilk}^\mathrm{Earth}} - \langle \ln A \rangle_{lk}^2.
    \label{eq:flux_mean_var}
\end{equation}
This precomputed database is then used in the implementation of the forward model in \texttt{stan} to perform Bayesian inference.

\subsection{Forward Modeling \& Inference with \texttt{stan}}
\noindent\textbf{Energy \& Mass Model: } 
The total UHECR energy spectrum at Earth is obtained by summing contributions from each injected mass species $A^S_l$ and all sources, including the diffuse background:
\begin{equation}
    \mathcal{J}(E) = \sum_{k=0}^{N_S}\sum_{l=1}^{N^A_\mathrm{inj}} f^A_{lk} F_k \mathcal{J}^\mathrm{Earth}_{lk}(E) = \sum_{k=0}^{N_S}\sum_{l=1}^{N^A_\mathrm{inj}} f^A_{lk} f^F_k \, F_\mathrm{tot} \mathcal{J}^\mathrm{Earth}_{lk}(E),
    \label{eq:energy_spectrum}
\end{equation}
where the expected number of detected events $N_k^\mathrm{ex}$ for each source is determined by weighting the detector exposure $\epsilon_T$ with the integrated energy spectrum. Here, the spectrum at Earth is weighted by the source- and composition-dependent mass fractions $f^A_{lk}$ and the source-dependent flux fractions $f^F_k$, multiplied by the total flux $F_\mathrm{tot}$. These weighted spectra are used to sample individual UHECR energies $E_\mathrm{Earth}$ in our inference procedure.

The mean and variance of $\ln A$ at Earth are similarly computed by summing over the injected mass species contributions, now weighting by the expected number fraction of events from each source, $N_k^\mathrm{ex}/N_\mathrm{tot}^\mathrm{ex}$, instead of the flux:
\begin{equation}
    \langle \ln A \rangle(E_\mathrm{bin}) = \sum_{k=0}^{N_S}\sum_{l=1}^{N^A_\mathrm{inj}} f^A_{lk} \frac{N_k^\mathrm{ex}}{N^\mathrm{ex}_\mathrm{tot}}\langle \ln A \rangle_{lk}, \quad
    \mathrm{Var}(\ln A)(E_\mathrm{bin}) = \sum_{k=0}^{N_S} \sum_{l=1}^{N^A_\mathrm{inj}} f_{lk}^A \frac{N_k^\mathrm{ex}}{N^\mathrm{ex}_\mathrm{tot}} \mathrm{Var}(\ln A)_{lk}.
    \label{eq:mean_var_lnA}
\end{equation}
Since event-by-event composition information is currently unavailable, we adopt a binned approach, sampling the mean and variance of $\ln A$ within discrete energy bins $E_\mathrm{bin}$ separate from the individually sampled energies in \Cref{eq:energy_spectrum}. Thus, mass composition measurements act as global external constraints within our Bayesian inference framework.
 \par 
\noindent\textbf{Detector Model: } 
UHECR measurements at observatories like Auger or TA are influenced by both statistical fluctuations and systematic uncertainties arising from calibration effects and hadronic model dependencies in the reconstruction process. To account for these effects, we model the detector response for each sampled energy $E$, and for each bin in $\langle \ln A \rangle$ and $\mathrm{Var}(\ln A)$, as truncated Gaussian distributions centered around the true values, including global systematic shifts:
\begin{align}
    E_\mathrm{det} &\sim \mathrm{TruncatedLogNormal}(\log E + \nu_{\log E},\, \sigma_{\log E},\, E_\mathrm{th},\, E_\mathrm{max}), \\
    \langle \ln A \rangle_\mathrm{det}(E_\mathrm{bin}) &\sim \mathrm{TruncatedNormal}(\langle \ln A \rangle(E_\mathrm{bin}) + \nu_{\mu \ln A},\, \sigma_{\mu \ln A}(E_\mathrm{bin}),\, 0,\, \infty), \\
    \mathrm{Var}(\ln A)_\mathrm{det}(E_\mathrm{bin}) &\sim \mathrm{TruncatedNormal}(\mathrm{Var}(\ln A)(E_\mathrm{bin}) + \nu_{\mathrm{Var} \ln A},\, \sigma_{\mathrm{Var} \ln A}(E_\mathrm{bin}),\, -1,\, \infty),
\end{align}
where $\nu_{\log E}$, $\nu_{\mu \ln A}$, and $\nu_{\mathrm{Var}\ln A}$ represent systematic uncertainties in the reconstructed energy, mean $\ln A$, and variance of $\ln A$, respectively. The corresponding statistical uncertainties, denoted by $\sigma_{\log E}$, $\sigma_{\mu \ln A}$, and $\sigma_{\mathrm{Var}\ln A}$, are allowed to vary within each energy bin. The truncation ensures physically meaningful results. Specifically, for the variance, a lower truncation bound of $-1$ is chosen to account for systematic shifts potentially leading to negative observed variance values.

\noindent\textbf{Prior Model: } 
We specify weakly informative priors for the model parameters as follows:
\begin{equation}
    \alpha_k \sim \mathcal{N}(-1,\, 3),\quad
    f^A_{k} \sim \mathrm{Dir}(\mathrm{Vec}(1,\, N^A_\mathrm{inj})),\quad
    f^F \sim \mathrm{Dir}(\mathrm{Vec}(1,\, N_S + 1)),\quad
    \log_{10} F_\mathrm{tot} \sim \mathcal{N}(-1,\, 3),
\end{equation}
where Dirichlet priors ensure that both the mass fractions $f^A_{k}$ and flux fractions $f^F$ sum to unity. 

\section{Results}

To verify our model, we apply it to simulated datasets based on the detector configurations of Auger and TA. We assume a single point source (now labeled SRC) located at $d = \SI{4}{\mega\parsec}$, motivated by potential source candidates (Centaurus A, M82) for both hemispheres at similar distances \cite{TelescopeArray:2018qwc,Yueksel2012}. The continuous injection from background sources (labeled BG) is also truncated at the same distance. We set identical source parameters for both detectors, specifically choosing: $\alpha_\mathrm{SRC} = -0.5$, $\alpha_\mathrm{BG} = 0.5$, $L_\mathrm{SRC} = \SI{7e41}{\erg\per\second}$, $f^A_\mathrm{SRC} = (0, 0.35, 0.5, 0.1, 0.05)$, and $f^A_\mathrm{BG} = (0, 0.2, 0.4, 0.3, 0.1)$. The source association fraction is set to $f_\mathrm{assos} = \sum_{k=1}^{N_S} N_k^\mathrm{ex} / N_\mathrm{tot}^\mathrm{ex} = 0.1$, defining the relative contribution of the point source(s) to the overall observed data. Simulated datasets are generated using \Cref{eq:energy_spectrum} to sample energies $E_\mathrm{Earth}$ for each UHECR, and \Cref{eq:mean_var_lnA} to sample the mean and variance of $\ln A$, utilizing energy bins from \cite{PierreAuger:2024nzw}. Detector responses are then applied to emulate observed energies and mass compositions, which we subsequently use in our inference model to reconstruct source parameters. \par 

\begin{table}[!h]
    \centering
    \begin{tabular}{|c|c|c|c|c|c|c|c|c|} \hline
        Detector & $N_\mathrm{ev}$ & $\epsilon_T$ / $\si{\kilo\meter\squared\year\steradian}$ & $\sigma_{\log E}$ & $\sigma_{\mu \ln A}$ & $\sigma_{\mathrm{Var} \ln A}$ & $\nu_{\log E}$ & $\nu_{\mu \ln A}$ & $\nu_{\mathrm{Var} \ln A}$ \\ \hline
        TA & 680 & 30500 & 0.2 & 0.2 & 0.2 & -0.05 & 0.3 & 0.5 \\
        Auger & 2750 & 122000 & 0.1 & 0.1 & 0.1 & 0.05 & 0.3 & 0.5 \\ \hline
    \end{tabular}
    \caption{Detector parameters used in this work. Statistical uncertainties reflect differences in detector exposure $\epsilon_T$, as indicated by the number of events $N_\mathrm{ev}$. Systematic uncertainties are motivated by the energy rescaling from the joint Auger-TA energy spectrum working group and upper limits from current $\ln A$ measurements.}
    \label{tab:detector_params}
\end{table}

\Cref{tab:detector_params} summarizes the statistical and systematic parameters for each detector model. To ensure comparability, we use the detector exposure from the Auger Phase One dataset \cite{PierreAuger:2022axr}, and take 1/4 of this total exposure for the detector exposure of TA. Correspondingly, statistical uncertainties for TA are doubled. Systematic uncertainties are derived from energy rescaling efforts by the Auger-TA energy spectrum working group \cite{TelescopeArray:2021zox} and upper limits from \cite{PierreAuger:2024nzw}. \par 

\begin{figure}[!h]
    \centering
    \includegraphics[width=0.44\linewidth]{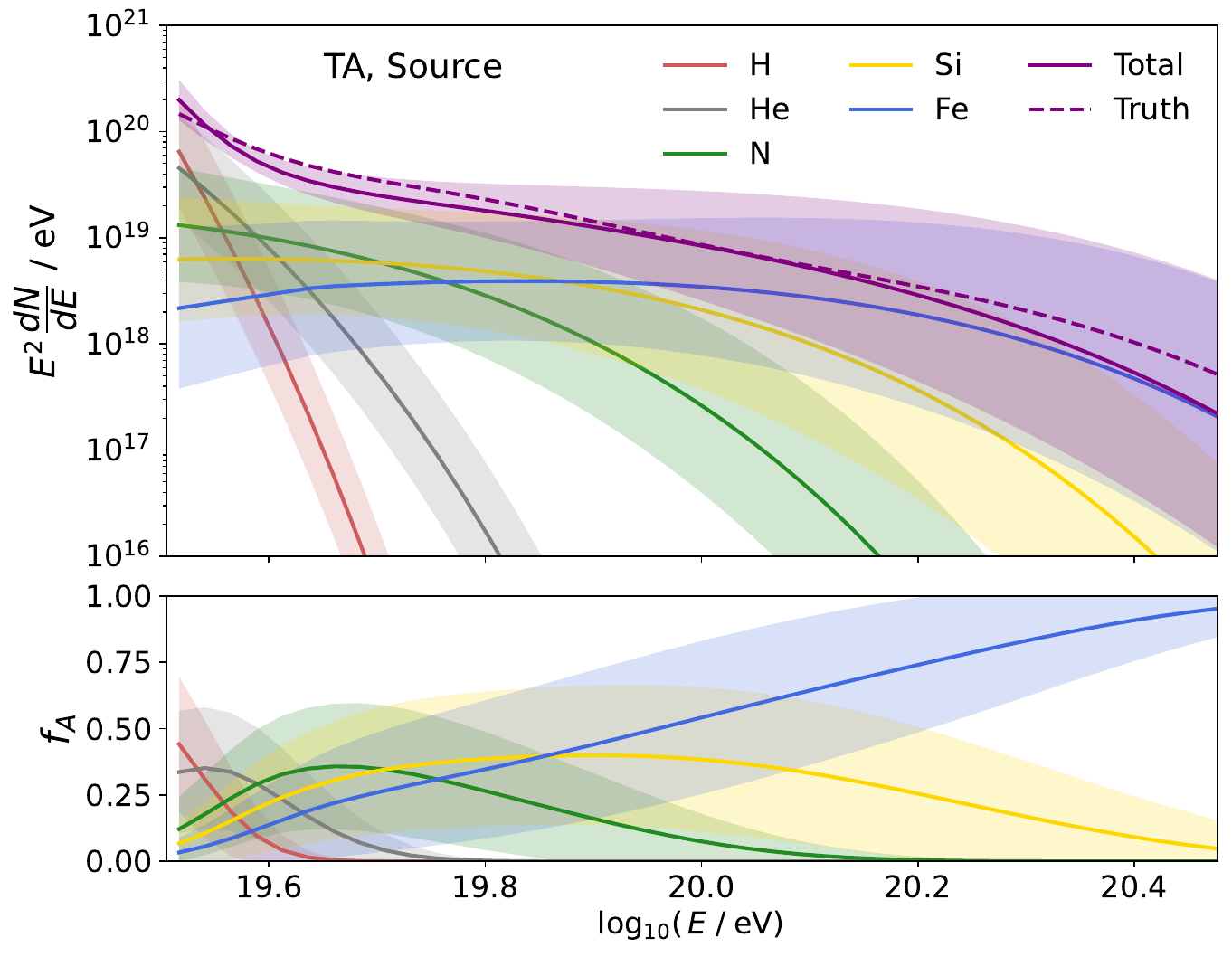} 
    \includegraphics[width=0.44\linewidth]{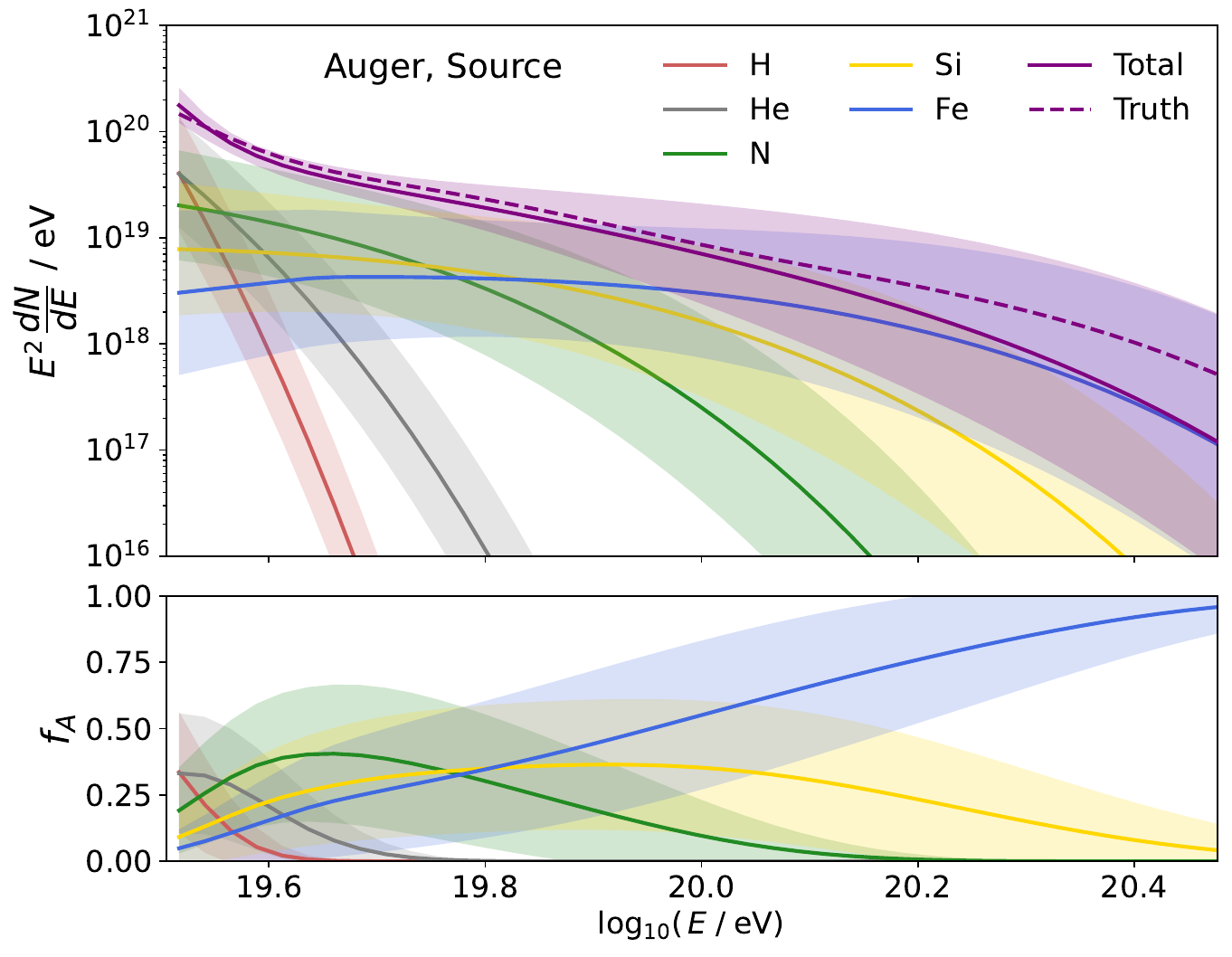} \\
    \includegraphics[width=0.44\linewidth]{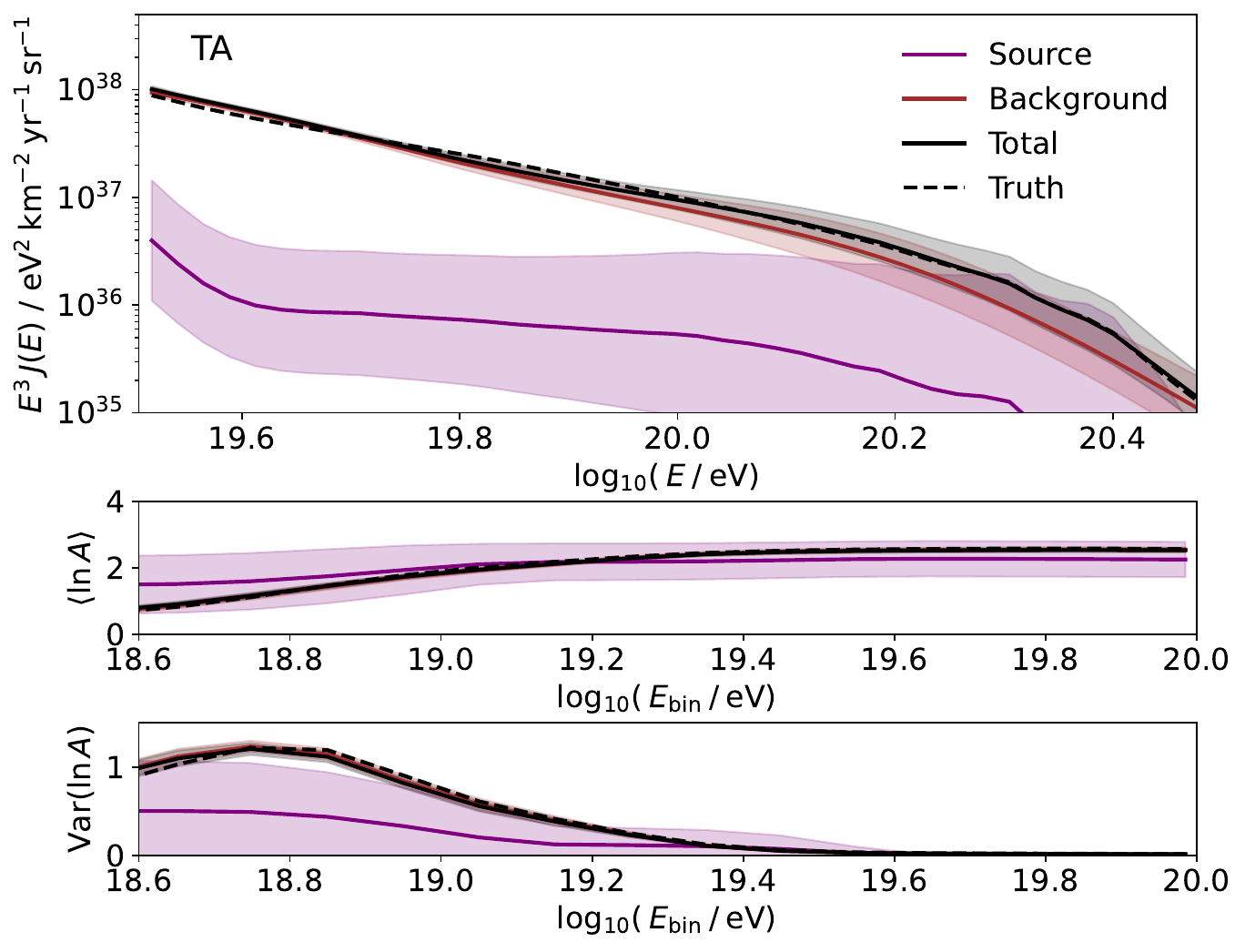} 
    \includegraphics[width=0.44\linewidth]{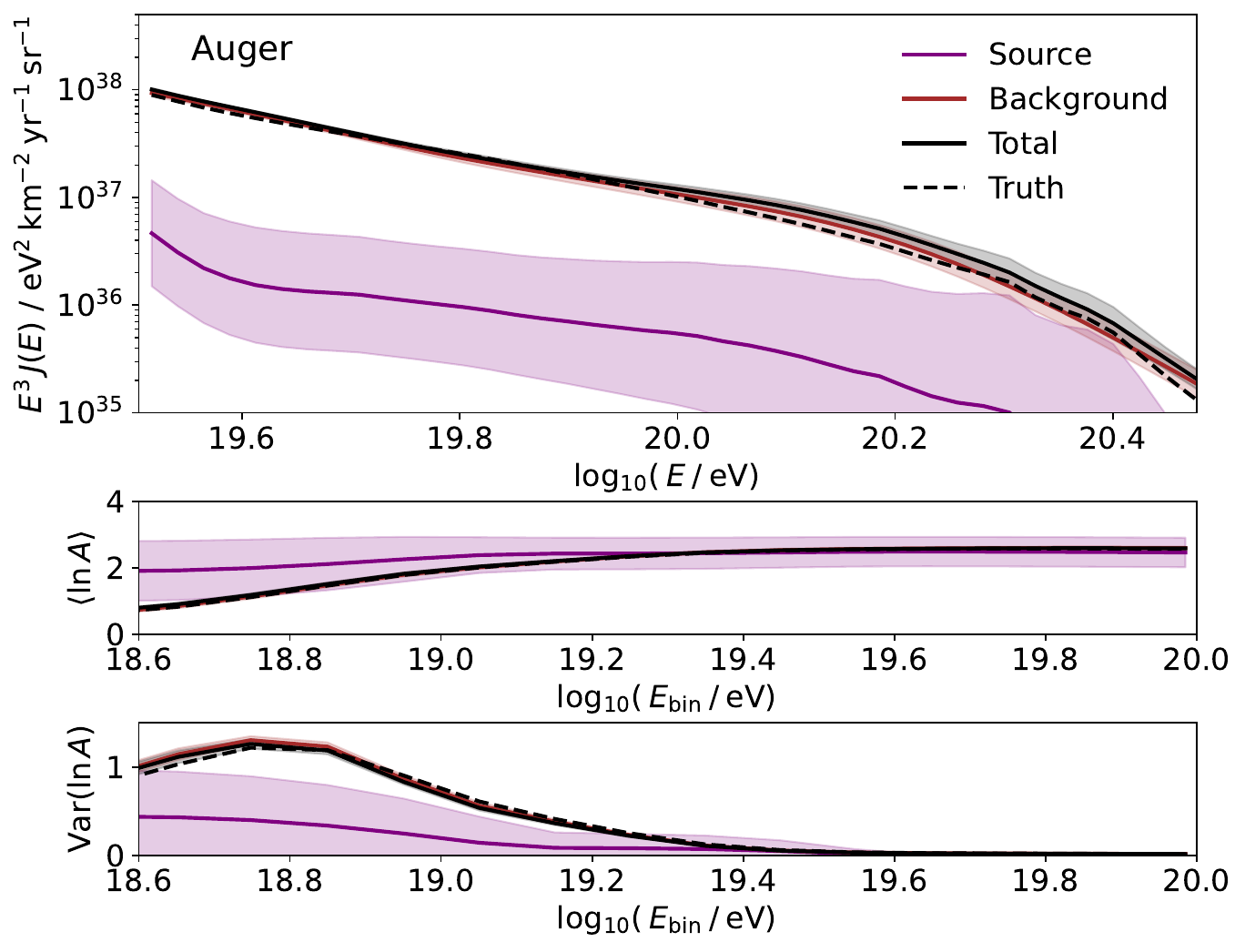} \\
    \includegraphics[width=0.44\linewidth]{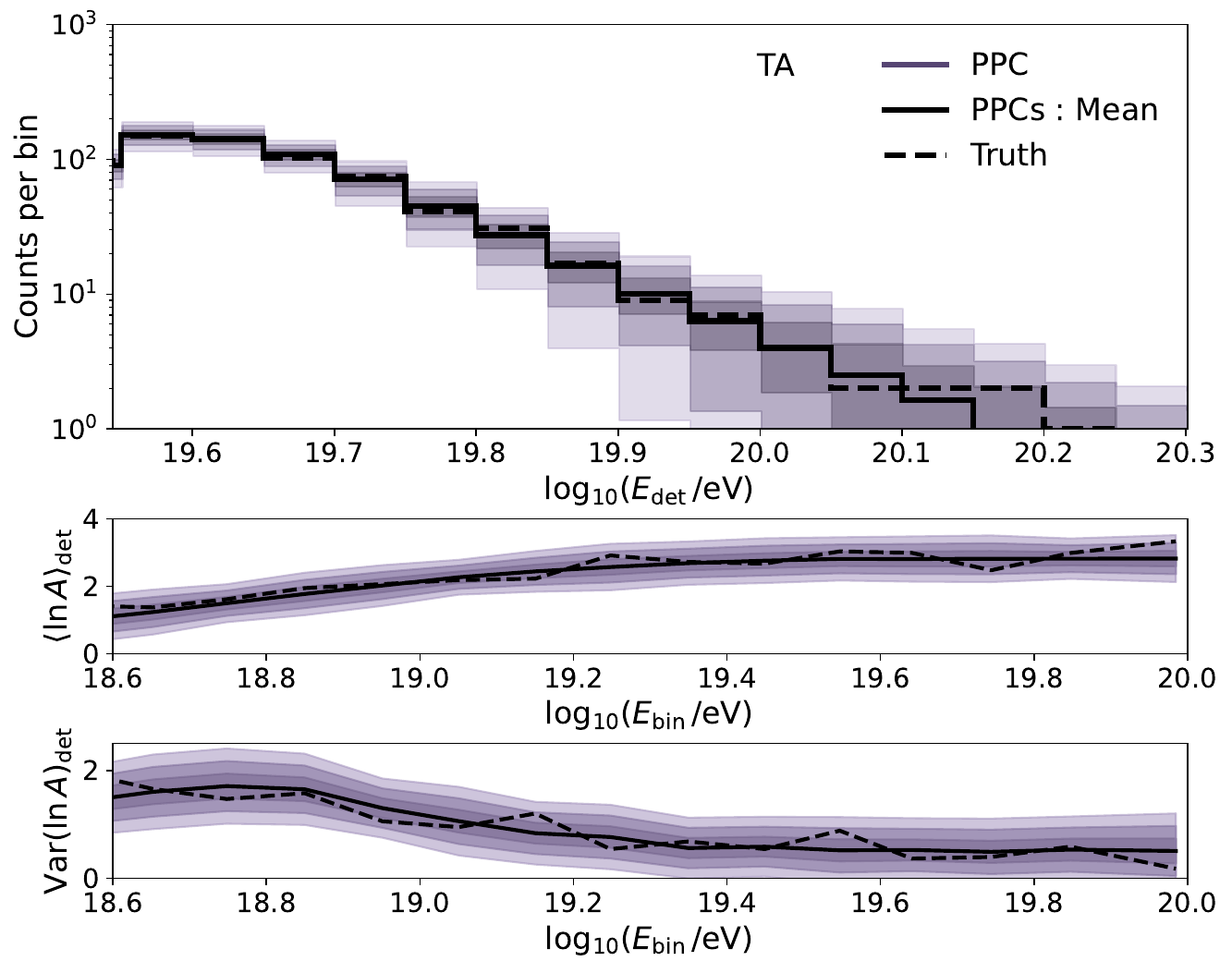}
    \includegraphics[width=0.44\linewidth]{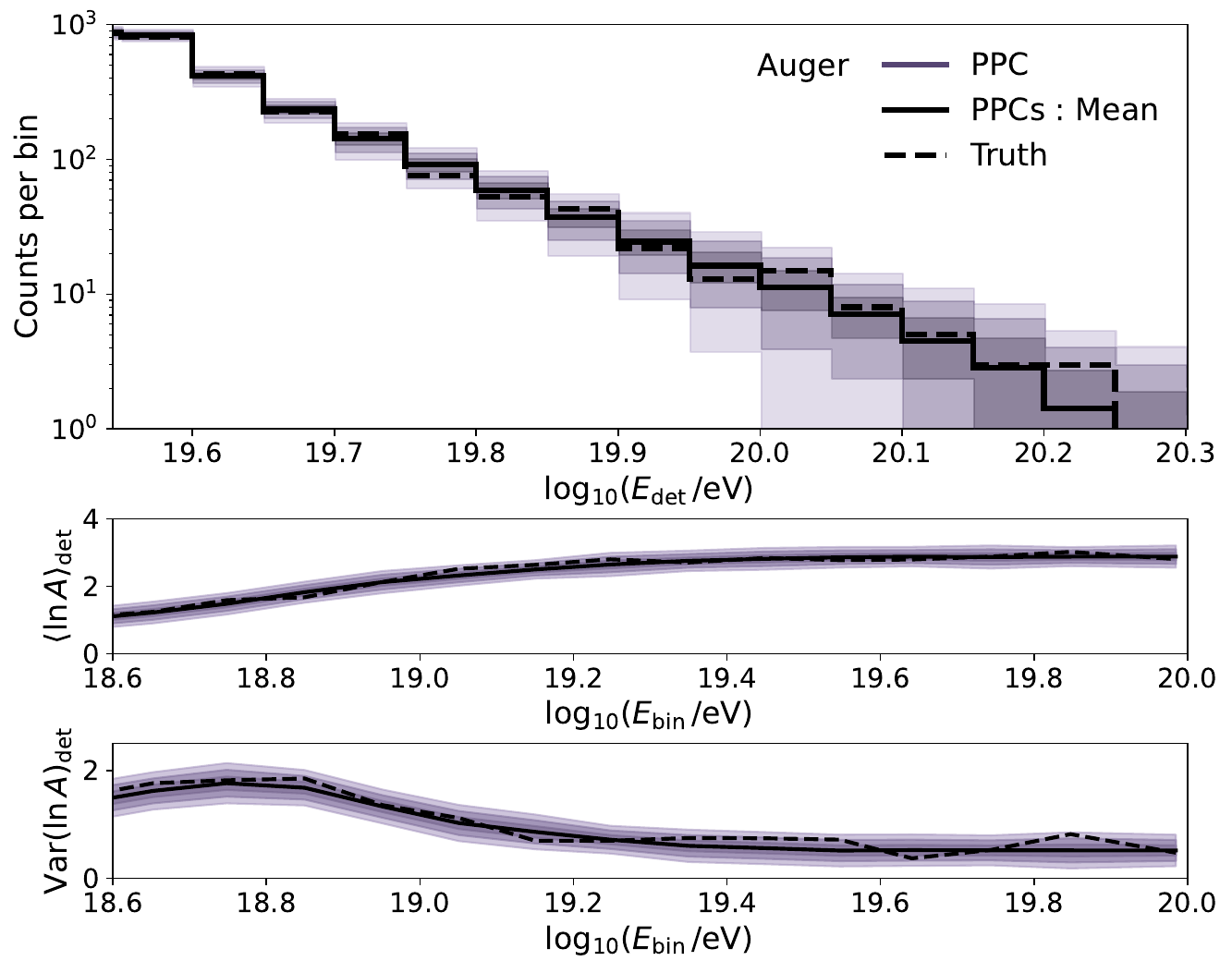}
    \caption{Reconstructed results from simulated datasets based on TA (left) and Auger (right). \textit{Top}: Inferred source spectrum of the point source, weighted by inferred mass fractions for each injected composition. Insets show spectrum-weighted mass fractions at each energy. \textit{Middle}: Inferred energy spectrum, weighted by the inferred flux contributions from the point and background sources. Insets display mean and variance of $\ln A$ per energy bin. \textit{Bottom}: Posterior predictive checks (PPCs), demonstrating forward-modeled results. Shaded regions represent the 1, 2, and 3$\sigma$ confidence intervals around the mean (black solid lines). True values are indicated by dashed lines.}
    \label{fig:results}
\end{figure}

\begin{figure}[!th]
    \centering
    \includegraphics[width=0.9\linewidth]{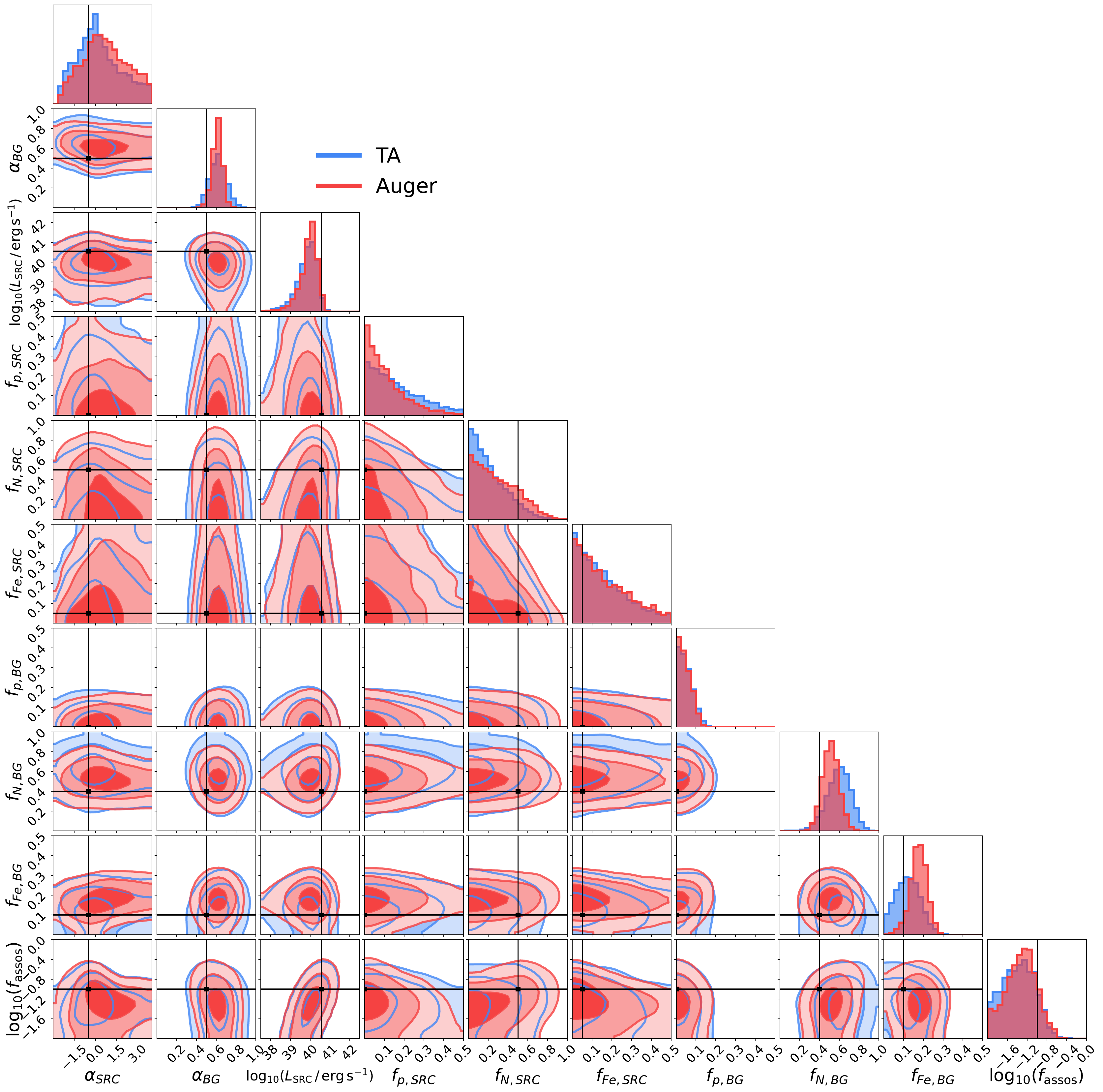}
    \caption{Corner plot of joint posterior distributions for key source parameters: spectral indices for point and background sources, source luminosity, proton, nitrogen, and iron mass fractions for point and background sources, and source association fraction. Results from TA (blue) and Auger (red) are shown, with contours representing 1, 2, and 3$\sigma$ intervals. True simulation values are indicated by black lines and dots. Figure generated using \texttt{corner.py} \cite{corner}.}
    \label{fig:corner}
\end{figure}

\Cref{fig:results} presents reconstruction results from simulated TA and Auger datasets. The inferred source spectra and spectrum-weighted mass fractions of each source component are recovered well, demonstrating accurate propagation of information through the model. The reconstructed energy spectra and mass composition moments for both detectors align well with true values, although the TA results exhibit larger uncertainties due to lower exposure. Nonetheless, the true parameter values consistently lie within the inferred 2$\sigma$ uncertainty intervals for all inferred source parameters, as shown in \Cref{fig:corner}.  \par 

We also perform posterior predictive checks (PPCs) by generating realizations of the simulated datasets using 100 posterior samples. This verifies the accurate reconstruction of detected energies within 1$\sigma$ confidence and the mass moments of $\ln A$ within 3$\sigma$, reflecting the reduced binning resolution employed in the current model. We also note that the inferred parameter contours in \Cref{fig:corner} are not exclusively dominated by statistical uncertainty; rather, they reflect limitations due to insufficiently detailed features in the spectral and average composition data alone. Therefore, incorporating spatial information and, ultimately, event-by-event mass measurements is expected to significantly improve constraints on source parameters.
\par

\section{Conclusion}

In this work, we presented a Bayesian hierarchical framework that integrates individual event energies and statistically averaged, binned mass composition data to infer the properties of UHECR sources, including luminosity, spectral indices, and injected mass fractions. The nuclear propagation code \texttt{PriNCe} is employed to model the observed energy spectrum and the mean and variance of $\ln A$ at Earth, clearly distinguishing between discrete injections from point sources and continuous injections from a homogeneous background. Our forward model also accounts for detector responses, incorporating both statistical and systematic uncertainties. Through application to simulated datasets representative of the Auger and TA observatories, we demonstrated that our framework reliably reconstructs the true source parameters at each stage of the modeling process.

Previously, we showed that rigidity-dependent deflections from Galactic and extragalactic magnetic fields can be incorporated alongside energy and mass composition models \cite{Watanabe:2023fdz,Watanabe:2023bmf,Watanabe:2025hsj}. Our future plans involve integrating this deflection model into the current framework, enabling a comprehensive analysis that utilizes all available observational information from existing UHECR observatories. We aim to apply this extended approach to publicly available data from Auger and TA, eventually expanding the analysis to include catalogs of potential point-source candidates such as starburst galaxies.


{\small
\paragraph{Acknowledgements} 
We acknowledge support from Academia Sinica under Grant No.~AS-GCS-113-M04. AF acknowledges additional support from the National Science and Technology Council under Grant No.~113-2112-M-001-060-MY3. Computational resources were provided by the Academia Sinica Grid Computing Center (ASGC) with support from the Institute of Physics, Academia Sinica.
}

\let\oldbibliography\thebibliography
\renewcommand{\thebibliography}[1]{%
  \oldbibliography{#1}%
  \setlength{\itemsep}{1pt}%
}

{\footnotesize
\bibliographystyle{JHEP}
\bibliography{publications}
}



\end{document}